\documentclass[sigconf, nonacm]{acmart}

\AtBeginDocument{%
  }

\setcopyright{acmlicensed}
\copyrightyear{2018}
\acmYear{2018}
\acmDOI{XXXXXXX.XXXXXXX}

\acmConference[Conference acronym 'XX]{Make sure to enter the correct
  conference title from your rights confirmation emai}{June 03--05,
  2018}{Woodstock, NY}
\acmISBN{978-1-4503-XXXX-X/18/06}

\begin{document}

\title{Equimetrics - Applying HAR principles to equestrian activities}

\author{Jonas Pöhler}
\email{jonas.poehler@uni-siegen.de}
\orcid{0000-0002-9942-8298}
\affiliation{%
  \institution{University of Siegen}
  \city{Siegen}
  \country{Germany}
}
\author{Kristof Van Laerhoven}
\orcid{0000-0001-5296-5347}
\affiliation{%
  \institution{University of Siegen}
  \city{Siegen}
  \country{Germany}
}

\renewcommand{\shortauthors}{Pöhler et al.}

\begin{abstract}
This paper presents the Equimetrics data capture system.
The primary objective is to apply HAR principles to enhance the understanding and optimization of equestrian performance. By integrating data from strategically placed sensors on the rider's body and the horse's limbs, the system provides a comprehensive view of their interactions. 
Preliminary data collection has demonstrated the system's ability to accurately classify various equestrian activities, such as walking, trotting, cantering, and jumping, while also detecting subtle changes in rider posture and horse movement. The system leverages open-source hardware and software to offer a cost-effective alternative to traditional motion capture technologies, making it accessible for researchers and trainers. The Equimetrics system represents a significant advancement in equestrian performance analysis, providing objective, data-driven insights that can be used to enhance training and competition outcomes.
The system has been made available at \cite{pohler_limlug/equimetrics:_2024}.
\end{abstract}


\begin{teaserfigure}
  \includegraphics[width=\textwidth]{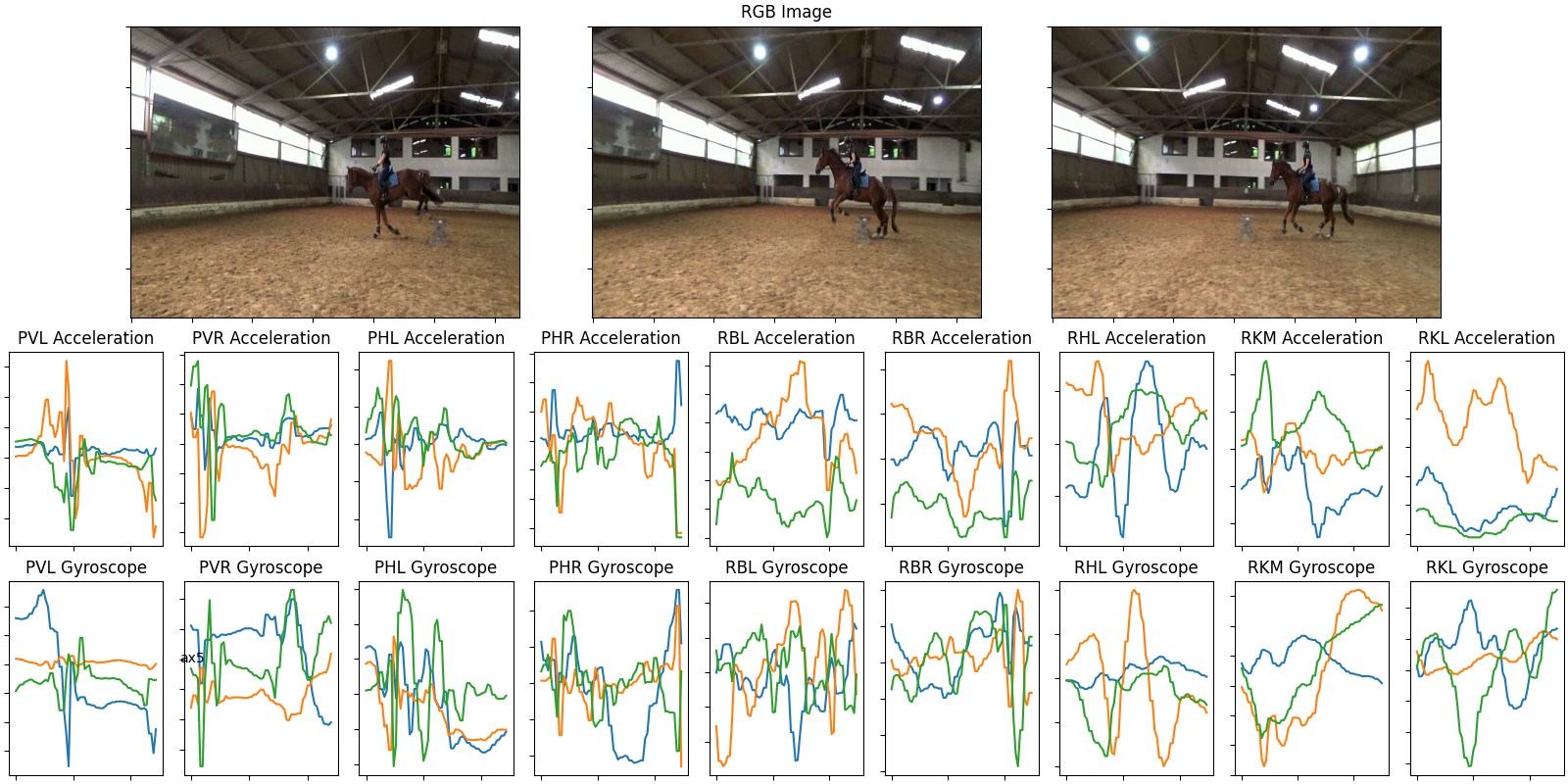}
  \caption{Exemplary captured data from the Equimetrics system during horse jumping, as recorded from 130 Hz Inertial Measurement Units that were positioned at the horses' ankles and the rider's wrist, ankle, waist, and head, all synchronized with external camera footage.}
  \label{fig:teaser}
\end{teaserfigure}

\received{20 February 2007}
\received[revised]{12 March 2009}
\received[accepted]{5 June 2009}

\maketitle

\section{Introduction}

The field of human activity recognition (HAR) has seen significant advancements in recent years, with researchers exploring the application of various sensor technologies and algorithms to recognize and classify human movements and behaviors  \cite{yu_liang_hsu_3c08a22b} \cite{chun_zhu_8354866a} \cite{zijuan_liu_d1f1d72b} \cite{jinghuan_guo_577f4836}.  One emerging area of interest is the application of these techniques to the domain of equestrian activities, where the complex interplay between rider and horse movements presents a unique challenge.

The Equimetrics sensor system is a novel approach to capturing and analyzing the motion of both the rider and the horse during equestrian activities.  This system utilizes a network of wearable inertial sensors to collect real-time data on the movements and interactions of the rider and horse, with the goal of applying human activity recognition principles to better understand and optimize equestrian performance  \cite{barbara_bruno_4cd8ab37} \cite{zijuan_liu_d1f1d72b} \cite{chun_zhu_8354866a} \cite{yu_liang_hsu_3c08a22b}. 

The use of wearable inertial sensors, such as accelerometers and gyroscopes, has been a common approach in human activity recognition research, as they can provide detailed information on the movements and orientations of the we arer's body  \cite{barbara_bruno_4cd8ab37} \cite{wesllen_sousa_lima_bc26586c}. These sensors can be strategically placed on the rider's body and the horse's limbs and torso to capture the complex biomechanics involved in equestrian activities. 

 This approach has been successful in recognizing human daily activities and sport activities, with researchers developing algorithms to classify motion patterns and detect transitions between different activities.  Similarly, the Equimetrics system aims to leverage these techniques to identify and differentiate between various equestrian activities, such as walking, trotting, cantering, and jumping, as well as to detect and analyze subtle changes in the rider's posture and the horse's movements that may be indicative of fatigue, strain, or other performance-relevant factors  \cite{chun_zhu_8354866a} \cite{yu_liang_hsu_3c08a22b} \cite{anniek_eerdekens_688c6b16} \cite{enrico_casella_23dcdebb} \cite{enrico_casella_75920176} \cite{jessica_maria_echterhoff_561a410f}.

By combining the data from the rider and horse sensors, the Equimetrics system can provide a comprehensive view of the equestrian interaction, enabling a deeper understanding of the factors that contribute to successful performance. This aligns with research on human and animal motion tracking using inertial sensors, which has highlighted the potential of such systems to provide valuable insights into the health and wellbeing of both humans and animals  \cite{f__marin_806a69c5}.

One key challenge in applying human activity recognition to equestrian activities is the need to address the complex interactions between the rider and the horse.  While previous research has primarily focused on single-person scenarios, the Equimetrics system must be able to recognize and differentiate between the movements and behaviors of both the rider and the horse, as well as their coordinated interactions. To address this, the system may incorporate additional sensors or computer vision techniques, as demonstrated in the DEEM system, which combined RFID and computer vision data to recognize exercises performed by multiple individuals in a gym environment  \cite{zijuan_liu_d1f1d72b}.
Another important consideration is the potential for variations in equestrian activities due to factors such as horse breed, size, and temperament, as well as the skill level and riding style of the rider.  The Equimetrics system must be able to adapt to these individual differences and provide accurate activity recognition across a diverse range of equestrian scenarios. 
Up until now these obsevations where only made through manual observation or with expensive and complex motion capture systems. \cite{kathryn_nankervis_e91b88c6} \cite{timo_de_waele_ad8d9580} \cite{anniek_eerdekens_6a776096}

Previously, the analysis of equestrian activities was largely reliant on subjective assessments of ideal movements by experienced trainers, based on their intuition and expertise. In contrast, the Equimetrics sensor system enables a more objective, data-driven approach to analyzing equestrian movements and training quality, by capturing and classifying the motion data from both the rider and the horse using wearable sensors.

\section{Related Work}
The application of human activity recognition principles to equestrian activities is a relatively new area of research, but there are several relevant studies that provide a foundation for the Equimetrics system.

One such study, "A framework for the recognition of horse gaits through wearable devices"  \cite{enrico_casella_23dcdebb}, developed a system for recognizing horse gaits using a smartphone and smartwatch.  The researchers placed sensors on the horse's saddle and the rider's wrist, and used machine learning algorithms to classify different horse gaits. 

Similarly, the "Smartwatch Application for Horse Gaits Activity Recognition" study also explored the use of wearable sensors to recognize horse gaits, with a focus on the effects of sliding window size and sampling frequency on the accuracy of the system.

These studies demonstrate the potential of wearable sensor technology to capture and analyze the complex movements involved in equestrian activities, and provide a starting point for the Equimetrics system. \cite{enrico_casella_75920176}

Additionally, the "Stochastic Recognition of Physical Activity and Healthcare Using Tri-Axial Inertial Wearable Sensors" study highlights the broader applications of human activity recognition in the healthcare and remote monitoring domains, which could be relevant to the Equimetrics system's goal of optimizing equestrian performance and monitoring the health and wellbeing of both the rider and the horse.

In many of the studies done in literature  \cite{enrico_casella_23dcdebb} \cite{kathryn_nankervis_e91b88c6}, the focus has been on using wearable sensors to recognize equine gaits and movements. However, the Equimetrics system aims to go beyond this by also considering the interactions between the rider and the horse, and how these interactions can be used to improve equestrian performance and overall well-being.

One study that touches on this aspect is the "Time-Series-Based Feature Selection and Clustering for Equine Activity Recognition Using Accelerometer Data" paper, which proposed a data-efficient algorithm for recognizing equine activities that considers both the horse and rider movements. \cite{timo_de_waele_ad8d9580}The researchers used a combination of feature selection and clustering techniques to develop a model that required only a small amount of labeled data, suggesting the potential for more nuanced activity recognition in the equestrian domain.

In the realm of human activity recognition IMU sensors are often used to classify motion patterns and detect transitions between different activities. \cite{timo_de_waele_ad8d9580} Especially for sport activities there is a lot of literature. 

Human activity recognition research has explored the potential of wearable inertial sensors to track and analyze complex movements, with applications in healthcare, sports, and other domains \cite{wesllen_sousa_lima_bc26586c} \cite{athanasios_i__kyritsis_f2405297} \cite{sara_garc_a_de_villa_1c67c27d} \cite{irvin_hussein_l_pez_nava_3241deae}.  These studies have demonstrated the ability of these sensors to capture detailed information on the orientation and motion of the wearer's body, and to use machine learning algorithms to classify different activities and detect transitions between them.
\section{Equimetrics Sensor System}
\begin{figure}[htp]
    \centering
    \includegraphics[width=0.89\linewidth]{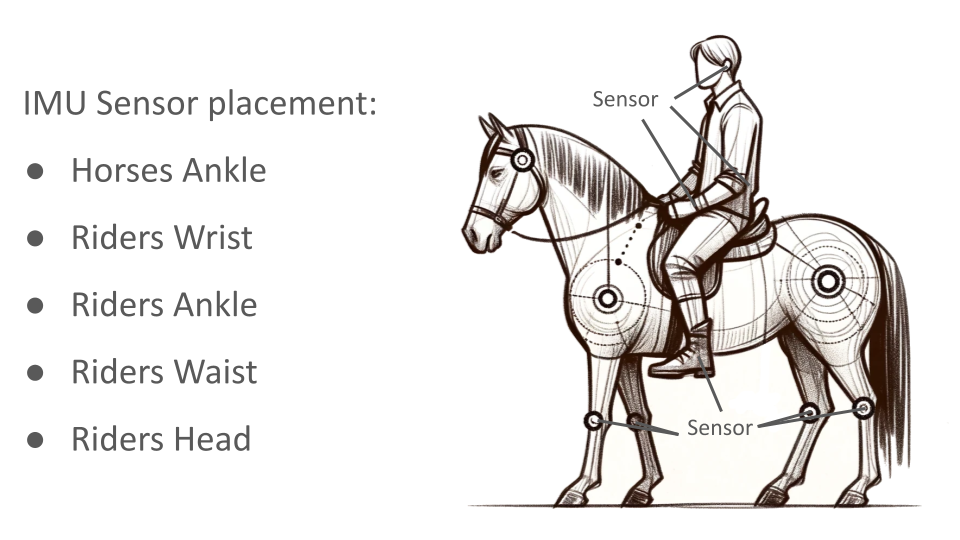}
    \caption{Positions of the IMU sensors.}
    \label{fig:sensor_placement}
\end{figure}
The Equimetrics sensor system consists of a network of 10 inertial measurement unit (IMU) sensors strategically placed on both the rider and the horse to capture comprehensive motion data from both sources. The sensors are installed on the rider's torso, head, arms, and legs, as well as on the horse's legs (see Figure \ref{fig:sensor_placement}), enabling a holistic view of the equestrian interaction. The sensor data is transmitted wirelessly to a central processing unit, which stores the information and is then able to recognize and classify the rider's and the horse's activities in later stages of the analysis.

Each sensor node in the system comprises a MPU-6050 NEMS with a 3-axis gyroscope and a 3-axis accelerometer, as well as an ESP32 microcontroller used to stream the data via Wi-Fi using a UDP data transfer protocol. In addition to the IMU sensors, the Equimetrics system also incorporates a video camera that tracks the rider's movements using the PIXEM camera system. This visual data enables a close observation of the rider's activities and can be used for labeling purposes, as well as for more detailed movement analysis with the OpenPose framework.

The central processing unit captures the 3D motion data from all the sensors at a high sampling rate of 130Hz. The accelerometer has a data range of ±16g, while the gyroscope has a range of ±2000 dps. The collected data is then preprocessed to ensure consistency across the sensor data, including sensor alignment, calibration, and resampling.

The Equimetrics system offers a cost-effective alternative to traditional motion capture technologies by leveraging open-source hardware and software. The hardware designs as well as the software are published openly, making the Equimetrics system accessible and affordable for researchers and equestrian trainers (published at anonymized). This open approach enables the creation of a multimodal dataset, combining motion data from the sensor network and visual data from the video cameras, to support the comprehensive analysis of equestrian activities.

Drawing on the insights and approaches from previous studies on human activity recognition and equestrian activity analysis, the Equimetrics system utilizes a combination of feature engineering and advanced machine learning techniques to recognize and classify the complex equestrian activities captured by the sensor network.
\section{Preliminary Data Capture Results}
To validate the capabilities of the Equimetrics sensor system, preliminary data collection and analysis was conducted on a small sample of riders and horses. The dataset consits of 45 minutes of data from two horses where the horses perform all the basic gaits as well as jumping over small obstacles. 

The data from the inertial sensors and the video cameras was synchronized and annotated to create a labeled dataset for activity recognition. 

The preliminary data capture demonstrated that the sensors maintained consistent synchronization and stability throughout the data collection period.
\section{Automatic Analysis}
\begin{figure}[htp]
    \centering
    \includegraphics[width=0.99\linewidth]{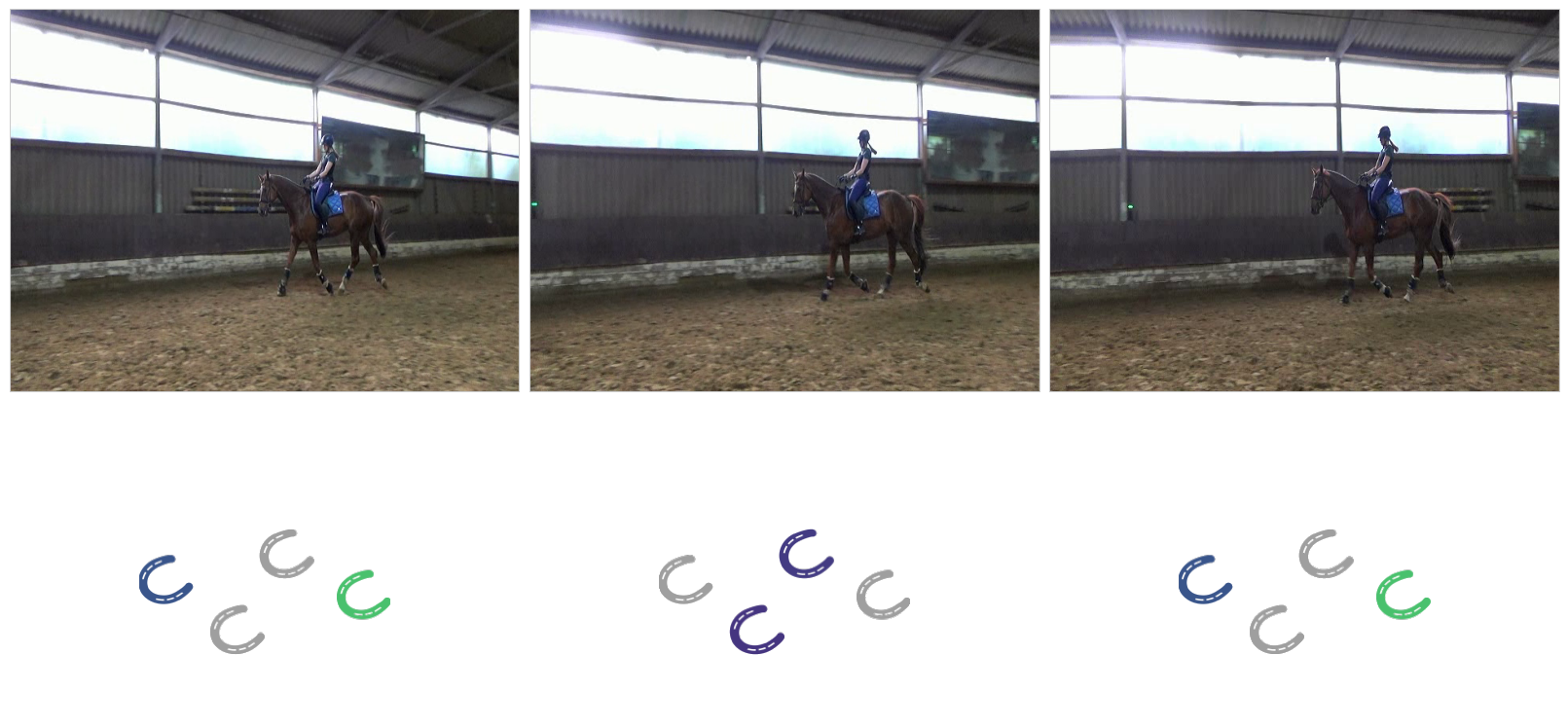}
    \caption{Highlighting the hoof-on and hoof-off events}
    \label{fig:hoof_placement}
\end{figure}
The high-frequency sensor data from the Equimetrics system offers substantial opportunities for automated analysis of equestrian activities. An initial step in the data analysis could involve leveraging the data from the four limb sensors on the horse to detect the precise timing of each hoof's contact with the ground. This could be achieved by fusing the accelerometer and gyroscope data from these sensors and converting it to quaternion representations, which would enable the identification of the individual hoof placement events. This detailed information on the horse's gait and movement patterns could provide valuable insights into the horse's performance and health, and could be used to optimize the rider's techniques and the overall equestrian interaction  \cite{m__tijssen_998195ac}.
The results presented in Figure \ref{fig:hoof_placement} demonstrate the effectiveness of the sensor data analysis for detecting the precise timing of hoof-on and hoof-off events. This was validated through comparison with annotated video data, which confirmed a high level of precision, with an average precision of 8.98 milliseconds. This performance aligns with the findings reported in the existing literature and represents a modest improvement over the results reported by Tjissen et al. The ability to accurately detect individual hoof placement events from the sensor data provides valuable insights into the horse's gait patterns and movement characteristics, which can be used to optimize the rider's techniques and assess the horse's overall performance and health.

\begin{figure}[htp]
    \centering
    \includegraphics[width=0.6\linewidth]{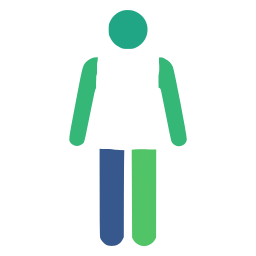}
    \caption{The extracted movements enables the computation of activity maps for the rider}
    \label{fig:activity_map}
\end{figure}

Furthermore, the combination of the rider and horse sensor data could enable the recognition of more complex equestrian activities, such as specific dressage movements, jumping techniques, or even the detection of rider falls or other safety-critical events. 
\begin{figure}[htp]
    \centering
    \includegraphics[width=0.6\linewidth]{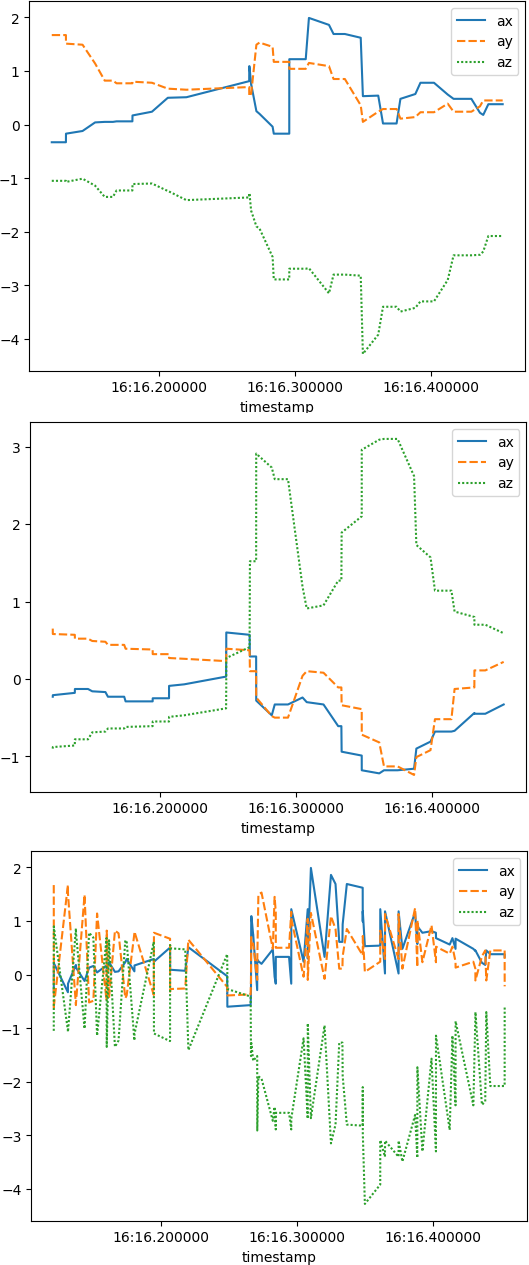}
    \caption{Extraction of the principal movement of the riders leg by substraction of horse movement}
    \label{fig:substraction}
\end{figure}
To discern the distinct movements of the horse's limbs and the rider's limbs individually is crucial. In equestrian activities, the sensor data from the rider captures a combination of the horse's movements and the rider's own movements. It is now possible to extract the rider's independent movement by subtracting the horse's movement from the combined data. The sensor placed on the rider's hips is particularly useful for this purpose, as it represents the point of greatest stability between the rider and the horse, resulting in the least amount of the rider's own movement and the highest overlap with the horse's movement. 
The data shown in Figure \ref{fig:substraction} represents the rider's leg movement with the horse's movement filtered out. This allows for a detailed assessment of the rider's independent limb movement, which can then be used to calculate a comprehensive movement magnitude index (MMI) that precisely quantifies and rates the specific characteristics of the rider's movements, as depicted in Figure \ref{fig:activity_map}. 
The MMI allows for the characterization of the rider's movement quality by quantifying the rider's ability to suppress unwanted motion. Additionally, it can be used to analyze the temporal relationship between the rider's movements and the horse's responses.
\section{Activity Recognition and Classification}
For the activity recognition component, a set of standardized equestrian test movements published by the Fédération Équestre Nationale (FN) was used as the basis for the analysis. This test protocol involved a combination of basic gaits, such as walk, trot, and canter, performed by the horses. The collected sensor data was then used to train two distinct Transformer-based human activity recognition models \cite{s22051911}. The first model focused on recognizing the specific gait of the horse, while the second model aimed to identify the dressage-related tasks being performed, such as specific movements and maneuvers. The test protocol consisted of 6 distinct movement tasks, some of which were executed in both directions to capture the horses' responses in different orientations. Due to the relatively short duration of certain tasks, 5-second data windows were utilized for the analysis to ensure the capture of the key movement characteristics. The Transformer models were trained on data from two different horses, each performing the test protocol twice, to increase the diversity of the training dataset and improve the models' generalization capabilities. The resulting F1 scores demonstrated a high level of accuracy, with 0.9324 for correctly identifying the horse's gait and 0.7601 for recognizing the specific dressage movements, indicating the effectiveness of the Equimetrics system in capturing and classifying equestrian activities.
\begin{figure}[htp]
    \centering
    \includegraphics[width=0.8\linewidth]{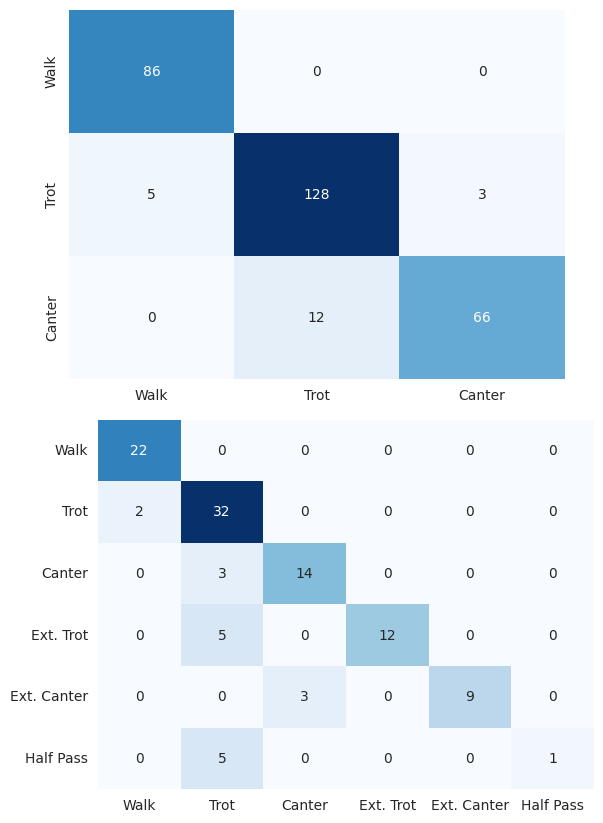}
    \caption{Confusion Matrix for the simple as well as for the complex classifier}
    \label{fig:confusion}
\end{figure}
The confusion matrices (see Figure \ref{fig:confusion}) for both classifiers demonstrate that the classification of the horse's gaits is reliable. However, in the case of the more complex dressage movements, the Half Pass movement exhibits a lower classification accuracy. This is likely due to the unique diagonal movement of the horse during the Half Pass, which differs from the other tested movements. Additionally, the limited number of samples for this specific movement may have contributed to the decreased classification performance.
\section{Discussion}
The Equimetrics sensor system has shown promising results in its ability to capture and analyze the movement patterns of both horse and rider. By leveraging advanced data analysis techniques, such as the precise detection of hoof placement events and the extraction of the rider's independent movement, the system provides valuable insights into the complex interactions between the rider and the horse. The activity recognition models, based on Transformer architectures, have demonstrated a high level of accuracy in identifying both the horse's gaits and the specific dressage-related movements, showcasing the potential of the Equimetrics system to support the evaluation and optimization of equestrian performance.

The integration of data from the horse and rider sensors facilitates a more holistic comprehension of the overall equestrian activity. By distinguishing the rider's independent movements from the horse's movements, the system can offer valuable feedback to the rider regarding their technique and coordination, thereby enabling the optimization of the rider-horse interaction. However, some limitations are associated with these preliminary findings, as the small number of horse-rider pairs examined constrains the generalizability of the results to a broader range of riders and horses. Further empirical investigations and structured data collection are necessary to address this limitation and expand the scope of the study. Evaluating the Equimetrics system with a larger and more diverse sample of horse-rider pairs would help validate the robustness of the activity recognition models and provide deeper insights into the nuances of equestrian performance. Additionally, longitudinal studies that track the progress of riders over time could shed light on the system's ability to facilitate long-term skill development and adaptation. Exploring these avenues would strengthen the evidence supporting the Equimetrics system as a valuable tool for enhancing equestrian training and competition.
\bibliographystyle{ACM-Reference-Format}
\bibliography{sample-base}

\end{document}